\documentclass[twocolumn,english,american,prl, singlespace, twocolumn]{revtex4-1}
\usepackage[T1]{fontenc}
\usepackage[utf8]{inputenc}
\usepackage{color}
\usepackage{bm}
\usepackage{amsmath}
\usepackage{amssymb}
\usepackage{graphicx}
\usepackage{esint}
\usepackage{babel}
\begin{document}
\title{Physical currents for stochastic Einstein-Podolsky-Rosen quantum trajectories}
\author{R. Y. Teh, M. Thenabadu, P.D. Drummond, M. D. Reid}
\affiliation{Centre for Quantum Science and Technology Theory, Swinburne University
of Technology, Melbourne 3122, Australia}
\begin{abstract}
Theories of the measured homodyne current generated by a stochastic
Schrödinger equation (SSE) can be tested in a simulation of the Einstein-Podolsky-Rosen
(EPR) correlations for a two-mode squeezed state. We carry out such
a simulation, and determine the correct stochastic term for the measured
current in the broad-band limit. Stratonovich rather than Ito stochastic
noise agrees with experiment. We show that this is relevant to measurement
noise and errors in quantum technologies. By analyzing the SSE trajectories
as measurement settings are changed, we propose a modern version of Schrodinger's
gedanken experiment, where one measures position and momenta simultaneously,
``one by direct, the other by indirect measurement''.
\end{abstract}
\maketitle
Trajectories generated by the stochastic Schrödinger equation (SSE)
\citep{diosi1989models,SSE} were originally proposed as models for
state reduction in quantum measurement. They are now used to simulate
the homodyne output measurements of quantum experiments \citep{Barchielli_1986,Gardiner1992wavefunction,belavkin1992nondemolition,Wiseman_PRA1993,Goetsch_PRA1994,Rigo_1997,Dorsselaer_2000,Gambetta_PRA2002,jacobs2006straightforward,barchielli2009quantum,wiseman2009quantum,carmichael2009statistical},
of increasing importance in quantum technology and foundational experiments.
Most applications to date have been limited to cases where the trajectory
modeling a detector current gives a realization of a single mode quantum
system \citep{carmichael2009statistical}. The value of the trajectory
is interpreted as providing a realization of the outcome of the measurement,
once quantum outputss are amplified.

In this Letter, we carry out SSE simulations of Einstein, Podolsky
and Rosen (EPR) correlations \citep{Einstein1935}. While EPR considered
spatially separated particles, realizations using homodyne measurements
on fields were later proposed and carried out experimentally \citep{ou1992realization,Reid:2009_RMP81,Reid:1989}.
 We simulate the moments measured in such an experiment with an SSE
that can model the output noise of the homodyne measurements.

Our analysis shows that the predicted current correlations change
drastically with the stochastic method used. Modeling the measurements
with Ito \citep{ito1951formula} stochastic noise gives incorrect
equal-time correlations either with or without a time-shift, because
only integrals of an Ito noise have physical meaning \citep{barchielli2009quantum}.
This can be resolved by including detector bandwidth effects \citep{carmichael2009statistical},
which leads to EPR correlations between time-integrated observables.
However, to obtain the correct unfiltered current satisfying an EPR
correlation (analogous to correlated positions $\hat{x}_{1}$ and
$\hat{x}_{2}$, and anti-correlated momenta $\hat{p}_{1}$ and $\hat{p}_{2}$)
we prove from adiabatic elimination that one must use a Stratonovich
\citep{stratonovich1971probability} noise at large bandwidth.

The correct choice of stochastic term is essential for understanding
wide-band continuous output measurements with multi-mode correlations.
SSE methods can simulate many quantum technology experiments, including
Bose-Einstein condensates \citep{wilson2007effects}, superconducting
quantum circuits \citep{minev2019catch}, the coherent Ising machine
(CIM) quantum computer \citep{Kewming2020quantum,Thenabadu_arxiv2025}
and, in future, LIGO gravitational wave detectors \citep{ma2017proposal}.
Large-scale quantum devices often utilize many modes and measurements
which display correlations and entanglement. As an example we apply
this to a strongly interacting CIM model used to solve NP-hard problems
\citep{McMahon_CIM_science2019}. We show that in the deep quantum
limit, changing the detector bandwidth alters errors due to shot noise,
which has a large effect on the success rate.

The SSE simulations also give insight into the assumptions behind
the EPR argument, which attempted to demonstrate the incompleteness
of quantum mechanics. EPR's argument was based on two premises: (1)
a criterion for an ``element of reality'' and (2) no action-at-a-distance.
For correlated spin states, the premises implied local hidden variable
descriptions which were subsequently negated by the work of Bell \citep{Bell1964,bell1966problem,Mermin1990-reality,Greenberger1989}.
However, EPR's assumptions do not break down at the macroscopic level
of the detector currents, where no-signaling is valid \citep{eberhard1989quantum}.
By analyzing the trajectories, as measurement settings $\theta$
are changed, we show consistency with a set of weakened EPR premises,
that apply to the currents and are not negated by Bell's theorem.
 As an illustration of EPR's argument, Schrödinger proposed one could
indirectly measure $\hat{p}_{1}$ of one particle, by measuring $\hat{p}_{2}$
of the correlated spatially-separated system, thereby simultaneously
measuring the position and momentum of a particle, ``one by direct,
the other by indirect measurement'' \citep{schrodinger1935gegenwartige,Colciaghi2023}.
We analyze the validity of this statement, by simulating Schrodinger's
gedanken experiment for the two-mode EPR fields, hence proposing
an experiment in which a wave-function measurement is interrupted
and changed in mid-collapse.

\paragraph*{Output field from the input-output relation}

We start by treating bosonic quadrature measurements on $M$ orthogonal
modes. The master equation for dissipation for a quantum density matrix
$\hat{\rho}$ with mode operators $\hat{a}_{k}$ and a Hamiltonian
$\hat{H}$ with units such that $\hbar=1$ is \citep{gardiner2004quantum,drummond_book}
\begin{align}
\frac{\partial\hat{\rho}}{\partial t} & =-i[\hat{H},\hat{\rho}]+\sum_{k=1}^{M}\gamma_{k}\left(\hat{a}_{k}\hat{\rho}\hat{a}_{k}^{\dagger}-\frac{1}{2}\left[\hat{n}_{k}\hat{\rho}+\hat{\rho}\hat{n}_{k}\right]\right)\,.\label{eq:Mast_eqn}
\end{align}
Here $\gamma_{k}$ is the number decay rate, the number operators
are $\hat{n}_{k}=\hat{a}_{k}^{\dagger}\hat{a}_{k}$, and we denote
operators as $\hat{O}$ to distinguish them from measured results.
The respective input and output fields $\hat{\bm{b}}_{in}$ and $\hat{\bm{b}}_{out}$
external to the bosonic system are related by the input-output relation
$\hat{\bm{b}}_{out}=\sqrt{\gamma}\hat{\bm{a}}+\hat{\bm{b}}_{in}$
\citep{Gardiner1985input}, where $\hat{\bm{a}}=(\hat{a}_{1},\hat{a}_{2})$.

For simplicity we suppose that the decay channels have equal damping
such that $\gamma_{k}=\gamma$, with dimensionless times $\tau=\gamma t$,
and dimensionless Hamiltonian $\tilde{H}=\hat{H}/\gamma$. We ignore
detector quantum efficiency, although this can be readily included.
 We treat two bosonic modes $k=1,2$ in a cavity or interferometer
that each decay to an external detector, with corresponding external
fields $\hat{b}_{k}$, which are also made dimensionless. 

In the case of a prepared photonic state in a vacuum, the quantum
stochastic operator equations can be solved exactly, using dimensionless
times and fields 
\begin{align}
\hat{\bm{b}}_{out}\left(\tau\right) & =e^{-\tau}\left(\bm{a}\left(\tau\right)+\intop_{0}^{\tau}e^{\tau'}\hat{\bm{b}}_{in}\left(\tau'\right)\,d\tau'\right)+\hat{\bm{b}}_{in}\left(\tau\right).
\end{align}
We define an internal vector quadrature operator $\hat{\bm{x}}=(\hat{x}_{1},\hat{x}_{2})=\hat{\bm{a}}+\hat{\bm{a}}^{\dagger}$,
with input and output quadrature operators $\hat{\mathbf{X}}_{in}=(\hat{X}_{in,1},\hat{X}_{in,2})$
and $\hat{\mathbf{X}}_{out}=(\hat{X}_{out,1},\hat{X}_{out,2})$, where
$\hat{\bm{X}}_{in(out)}=\hat{\bm{b}}_{in(out)}+\hat{\bm{b}}_{in(out)}^{\dagger}$.
Hence, $\hat{X}_{out,k}=\hat{x}_{k}+\hat{X}_{in,k}.$ If the input
field is in a vacuum state with $\langle\bm{X}_{in}(\tau)\rangle_{Q}=0,$
one gets mean values: $\langle\hat{\bm{X}}_{out}\rangle_{Q}=e^{-\tau/2}\langle\hat{\bm{x}}(0)\rangle_{Q}=\langle\hat{\bm{x}}(\tau)\rangle_{Q}\,,$where
$\langle.\rangle_{Q}$ is a quantum ensemble average.

The relationship of external field $\hat{X}_{out,k}$ and the measured
current operator $\hat{J}_{k}$ depends on the local oscillator and
the detector gain \citep{carmichael2009statistical}. After rescaling
to a dimensionless form, the ideal output current in the wide-band
limit is simply $\hat{\bm{J}}=\hat{\bm{X}}_{out}$ where $\hat{\mathbf{J}}=(\hat{J}_{1},\hat{J}_{2})$.
Since the measured current has a finite bandwidth, this can only hold
over a restricted bandwidth. 

To analyze EPR correlations, we generalize this to a balanced homodyne
scheme for measuring the internal quadrature $\hat{x}_{k}^{\phi_{k}}=\hat{x}_{k}\cos\phi_{k}+\hat{p}_{k}\sin\phi_{k}$,
by combining the output field    with a macroscopic local oscillator
(LO) field, each with an independent phase-shift $\phi_{k}$ for each
mode $k$. To treat this, we define $\tilde{a}_{k}=\hat{a}_{k}e^{-i\phi_{k}}$,
and the rotated quadrature as $\tilde{x}_{k}=\tilde{a}_{k}+\tilde{a}_{k}^{\dagger}$
, for measurements with a fixed local oscillator phase.

\paragraph*{Homodyne current using a stochastic equation}

There is an SSE equivalent to Eq (\ref{eq:Mast_eqn}). This scales
linearly in the Hilbert space dimension, giving a lower complexity
than the master equation. In its simplest form it is a stochastic
differential equation (SDE) \citep{carmichael1993quantum,Goetsch_PRA1994,gardiner2004handbook,wiseman2009quantum,carmichael2009statistical}
following Ito \citep{ito1951formula,Gardiner1997} calculus:

\begin{align}
\frac{d\left|\Psi\right\rangle _{I}}{d\tau} & =\left\{ -i\tilde{H}+\left(\left\langle \tilde{\bm{x}}\right\rangle _{I}-\tilde{\bm{a}}^{\dagger}\right)\cdot\frac{\tilde{\bm{a}}}{2}-\frac{\left\langle \tilde{\bm{x}}\right\rangle _{I}^{2}}{8}+\Delta\tilde{\bm{a}}\cdot\bm{\xi}\right\} \left|\Psi\right\rangle .\label{eq:Ito_SSE}
\end{align}
Here $\left|\Psi\right\rangle _{I}$ is a state conditioned on a pseudo-current
$\bm{\bm{j}^{I}=}(j_{1}^{I},j_{2}^{I})$ with an Ito noise $\bm{\xi}^{I}$,
the fluctuation operator is $\Delta\tilde{\bm{a}}\equiv\tilde{\bm{a}}-\frac{1}{2}\left\langle \tilde{\bm{x}}\right\rangle _{S}$
and
\begin{equation}
\bm{j}^{I}(\tau)=\left\langle \tilde{\bm{x}}(\tau)\right\rangle _{I}+\bm{\xi}^{I}\left(\tau\right).\label{eq:Ito_current}
\end{equation}
The real noise $\bm{\xi}^{I}$ is defined such that $\left\langle \xi_{k}^{I}\left(\tau\right)\xi_{j}^{I}\left(\tau'\right)\right\rangle =\delta_{kj}\delta\left(\tau-\tau'\right)$,
where $\xi_{k}^{I}$ is the noise for mode $k$, and $\left\langle \hat{\bm{x}}(\tau)\right\rangle _{I}=\left\langle \Psi\right|\tilde{\bm{x}}\left|\Psi\right\rangle _{I}$.

There are several proposals for interpreting $\bm{j}^{I}(\tau)$
as a realistic sample of measured output currents $\bm{J}$, whose
ensemble averages and correlations must match the quantum predictions
\citep{Wiseman_PRA1993,Patra2022,jacobs2006straightforward,gardiner2004quantum}.
Since Ito noise has unusual mathematical properties, it is important
to clarify this interpretation as it becomes more relevant to modern
quantum experiments.

We will show that an Ito interpretation gives no initial EPR correlations,
completely different to the quantum prediction. This is because Ito
calculus requires corrections \citep{ito1951formula} that do not
occur for correlations of physical measurements. Evaluating the Ito
noise at an earlier time to the wave-function, so that $\bm{j}^{d}=\left\langle \tilde{\bm{x}}(\tau)\right\rangle _{I}+\bm{\xi}^{I}\left(\tau-d\tau\right)$
\citep{Wiseman_PRA1993,Gambetta_PRA2002}, also gives incorrect correlations.

Another approach is to derive a Stratonovich SDE from (\ref{eq:Ito_SSE}).
This is the wide-band limit of a finite bandwidth stochastic equation
\citep{stratonovich1971probability} following standard calculus.
There are several forms \citep{Gambetta_PRA2002}, but we use an SSE
equivalent to (\ref{eq:Ito_SSE}), explained in Appendix A, giving:

\begin{equation}
\frac{d\left|\Psi\right\rangle _{S}}{d\tau}=\left(-i\tilde{H}+\Delta\tilde{\bm{a}}\cdot\bm{j^{S}}+\frac{\left\langle \tilde{\bm{x}}^{2}\right\rangle -M}{4}-\frac{\tilde{\bm{x}}\tilde{\bm{a}}}{2}\right)\left|\Psi\right\rangle _{S}.\label{eq:Strat_SSE}
\end{equation}
Here the fluctuation operator is $\Delta\tilde{\bm{a}}\equiv\tilde{\bm{a}}-\frac{1}{2}\left\langle \tilde{\bm{x}}\right\rangle _{S}$
and the Stratonovich pseudo-current $\bm{j}^{S}=(j_{1}^{S},j_{2}^{S})$
is
\begin{equation}
\bm{j}^{S}(\tau)=\left\langle \tilde{\bm{x}}(\tau)\right\rangle _{S}+\bm{\xi}^{S}(\tau).\label{eq:Stratonovich_noise}
\end{equation}
This is defined using Stratonovich noise $\bm{\xi}^{S}$ with the
same correlations as before. We show below that the physical wide-band
current is simply $\bm{J}(\tau)=\bm{j}^{S}(\tau)$, which gives correct
wide-band EPR correlations.

The proof is based on more rigorous characteristic functional methods,
explained in Appendix B, which show that only the time-integral of
an Ito noise is physical \citep{barchielli2009quantum}. To obtain
the physical current, we now assume a finite bandwidth detector to
give a second coupled SDE \citep{carmichael2009statistical} for the
detected photocurrent $\bm{J}=(J_{1},J_{2})$:
\begin{align}
\frac{d\bm{J}}{d\tau} & =-\kappa\left(\bm{J}-\bm{j}\right).
\end{align}
In this approach, the current $\bm{J}$ has a finite bandwidth $\kappa,$
and follows standard calculus. Since Ito equations can be transformed
to Stratonovich equations using known rules \citep{gardiner2004handbook,Arnold1992-stochastic},
this resolves the Ito vs. Stratonovich ambiguity. 

In the wide-band detector limit of $\kappa\rightarrow\infty$, one
can adiabatically eliminate the 'fast' variable $\bm{J}$, so that
$\dot{\bm{J}}=0$. This type of adiabatic elimination is only valid
in the case of a Stratonovich SDE \citep{gardiner1984adiabatic},
and gives that $\bm{J}\rightarrow\bm{j}^{S}$. Hence the physical
current in the limit of a wide-band detector is the Stratonovich current.

\paragraph*{Two-mode squeezed state}

To explain the physical EPR argument, consider two interferometers
prepared in a two-mode squeezed state, which is the most practical
route for implementing the quantum correlations proposed in the original
EPR gedanken-experiment \citep{Reid:1989}. The initial state is $|TMSS\rangle,$
defined as
\begin{align}
|TMSS\rangle & =\frac{1}{\cosh r}\sum_{n=0}^{N_{c}}(\text{tanh}r)^{n}{\color{red}{\color{black}|n\rangle_{1}|n\rangle_{2}}}\,,
\end{align}
where $r$ is the squeezing parameter, $|n\rangle_{k}$ is a number
state for mode $k$ and $N_{c}$ is the photon cutoff, taken as $N_{c}\rightarrow\infty$
in the idealized case.  The time evolution of the internal quantum
correlation $\langle\hat{x}_{1}\hat{x}_{2}\rangle_{Q}$ from the operator
equations has an analytical solution:
\begin{align}
\langle\hat{x}_{1}\left(\tau\right)\hat{x}_{2}\left(\tau\right)\rangle_{Q} & =e^{-\tau}\sinh(2r)\,
\end{align}
Similarly, defining $\hat{p}_{k}=(\hat{a}-\hat{a}^{\dagger})/i$,
one finds that $\langle\hat{p}_{1}\left(\tau\right)\hat{p}_{2}\left(\tau\right)\rangle_{Q}=-e^{-\tau}\sinh(2r)$,
and also, $\langle\hat{x}_{i}\left(\tau\right)\hat{x}_{i}\left(\tau\right)\rangle_{Q}=\langle\hat{p}_{i}\left(\tau\right)\hat{p}_{i}\left(\tau\right)\rangle_{Q}=e^{-\tau}\cosh(2r).$

\begin{figure}
\begin{centering}
\includegraphics[width=0.75\columnwidth]{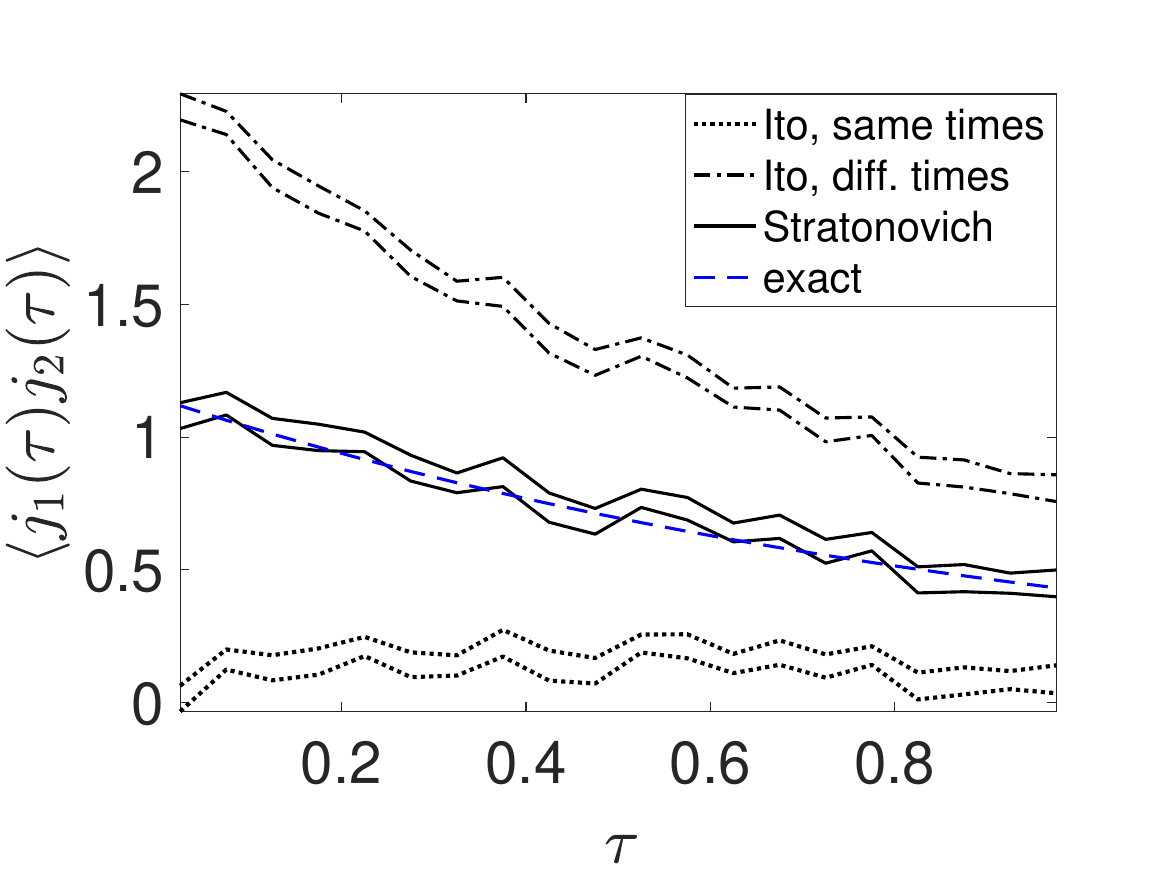}
\par\end{centering}
\caption{The averaged unfiltered SSE homodyne current correlation $\left\langle j_{1}(\tau)j_{2}(\tau)\right\rangle $
vs the dimensionless time $\tau=\gamma\tau$ for a two-mode damped
squeezed state, comparing different theories. The blue dashed line
is the infinite bandwidth quantum solution, $\langle\hat{J}_{1}(\tau)\hat{J}_{2}(\tau)\rangle=-e^{-\tau}\sinh(2r)$,
in agreement with the Stratonovich result (where $j_{k}\equiv j_{k}^{S}$),
but far from the two Ito results (where $j_{k}\equiv j_{k}^{I}$),
whether one evaluates the noise and the wave-function either at the
same time or not. Here, $r=0.5$, the two lines are $\pm\sigma$
sampling error bars for a sample size of $2\times10^{5}$, and the
time step-size is $0.05$. \label{fig:j1j2_unfiltered}}
\end{figure}

 For an SSE current $j_{k}$ to be valid, it should satisfy the same
correlations at equal times: i.e. 
\begin{equation}
\left\langle j_{1}(\tau)j_{2}(\tau)\right\rangle =\langle\hat{J}_{1}(\tau)\hat{J}_{2}(\tau)\rangle_{Q}.\label{eq:current-x-1}
\end{equation}
The correlation (\ref{eq:current-x-1}) is measurable and directly
reflects that between the two internal cavity modes, which leads to
an EPR paradox. We find $\langle(\hat{x}_{1}(0)-\hat{x}_{2}(0))^{2}\rangle\rightarrow0$
and $\langle(\hat{p}_{1}(0)+\hat{p}_{2}(0))^{2}\rangle\rightarrow0$
implying $\hat{x}_{1}(0)=\hat{x}_{2}(0)$ and $\hat{p}_{1}(0)=-\hat{p}_{2}(0)$,
as $r\rightarrow\infty$ .  The value of $\hat{x}_{2}(0)$ can be
predicted from $\hat{x}_{1}(0)$; the value of $\hat{p}_{2}(0)$ can
be predicted from $\hat{p}_{1}(0)$. The external fields are less
strongly correlated, but an inferred Heisenberg uncertainty relation
is violated for all $r$, thereby satisfying an EPR criterion \citep{Reid:1989,Reid:2009_RMP81}.
 We show in Appendix C and in the numerical results that the condition
(\ref{eq:current-x-1}) requires the Stratonovich SSE. Appendix D
shows that EPR correlations are not obtained  with an Ito SSE (Fig.
\ref{fig:j1j2_unfiltered}).  The Stratonovich SSE simulation confirms
that the values for the stochastic currents $j_{1}(\tau)$ and $j_{2}(\tau)$
are correlated in the manner expected for EPR correlations.

\paragraph*{Elements of reality at finite bandwidth}

 EPR's criterion of reality is that ``if, without in any way disturbing
a system, we can predict with certainty the value of a physical quantity,
then there exists an element of reality corresponding to this quantity''.
 The instantaneous current $J_{k}(\tau)$ gives a measure of the
external amplified quadratures, $\hat{X}_{k}^{out}$ or $\hat{P}_{k}^{out}$,
but contains extra fluctuations due to the input vacuum fields. While
this vanishes in the correlation $\langle J_{1}(\tau)J_{2}(\tau)\rangle$,
the noise manifests in the variances $\langle(\hat{X}_{1}^{out}(\tau)-\hat{X}_{2}^{out}(\tau))^{2}\rangle/$
and $\langle(\hat{P}_{1}^{out}(\tau)+\hat{P}_{2}^{out}(\tau))^{2}\rangle$,
and hence in the predictions for $\hat{X}_{\theta_{2},2}^{out}$ given
measurement of $\hat{X}_{\theta_{1},1}^{out}$. To account for this
\citep{drummond1989time}, we can consider a sequence of the time
averaged current outputs over intervals $[\tau_{n}^{-},\tau_{n}^{+}]$
where $\tau_{n}^{\pm}=\left(n\pm\frac{1}{2}\right)\Delta\tau$, and
$J_{k,n}=\int_{\tau_{n}^{-}}^{\tau_{n}^{+}}J_{k}\left(\tau\right)d\tau$,
which measures the time-averaged observable $\tilde{x}_{k,n}=\frac{1}{\sqrt{\Delta\tau}}\int_{\tau_{n}^{-}}^{\tau_{n}^{+}}\tilde{X}_{k}\left(\tau\right)d\tau$,
for each external system, $k=1,2$. The value $J_{k,1}$ can be predicted
by a measurement of $J_{k,2}$ to better than the inferred Heisenberg
uncertainty principle \citep{Reid:1989}, following EPR's criterion
for an ``element of reality''. A detailed treatment of time averaging
is given in Appendix E.

Time-averaging removes the difference between the Ito and Stratonovich
predictions, but without time-averaging, the equal-time correlations
are not identical. 

\paragraph*{Schrodinger's gedanken experiment}

We now investigate the nature of the element of reality (EOR) in the
SSE simulation. By choosing settings $\theta_{1}=0$ and $\theta_{2}=\pi/2$,
we construct a realization of Schrodinger's proposal to measure both
$\hat{x}$ and $\hat{p}$ simultaneously, by measurement of $\hat{p}_{2}$
at system $2$ and $\hat{x}_{1}$ at system $1$. We define a time
$t_{D2}$, when the stochastic current $J_{2}(\tau)$ at system $2$
has been generated, but prior to photo-detection at system $1$. The
value of the stochastic current $J_{2,av}(\tau)$ implies an outcome
$p_{2}$ for $\hat{p}_{2}$.
\begin{figure}[t]
\begin{centering}
\includegraphics[width=0.9\columnwidth]{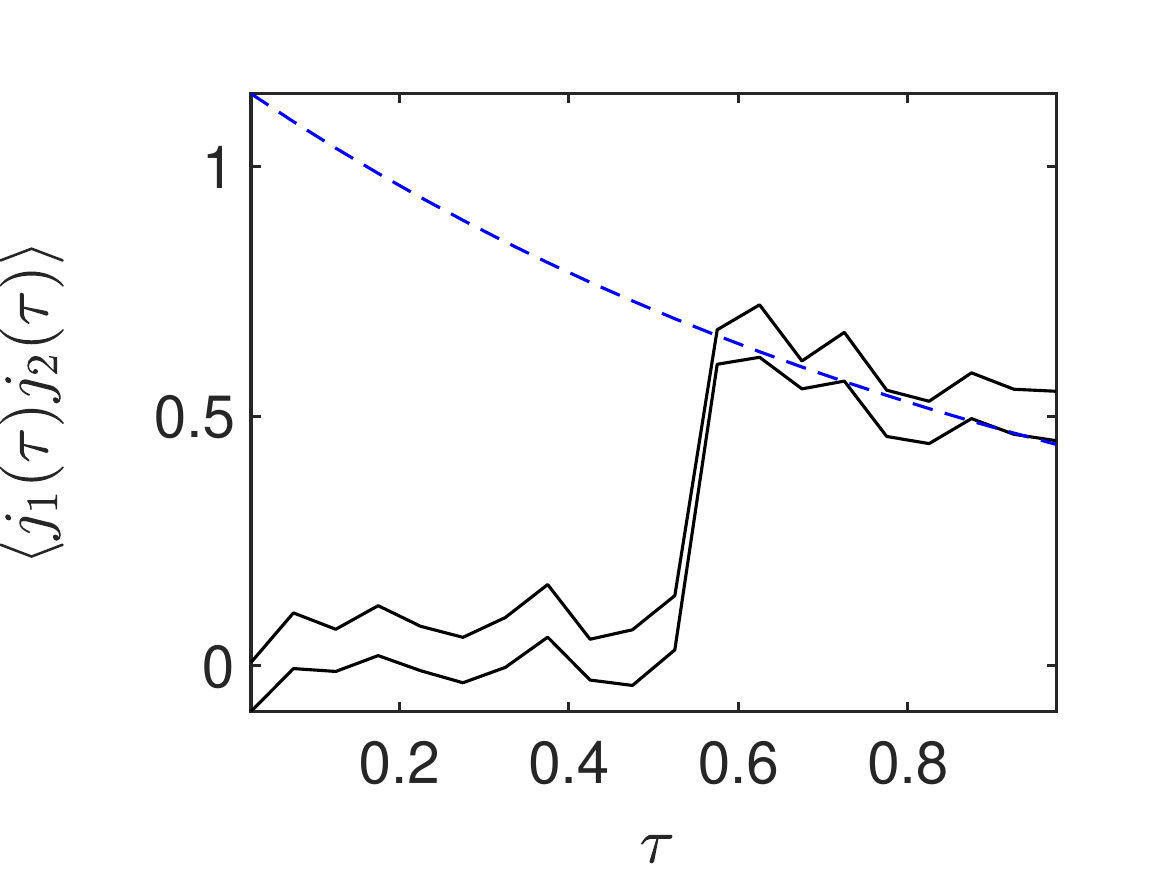}
\par\end{centering}
\caption{\foreignlanguage{english}{Graph of correlations when one phase angle is changed dynamically,
so that at first the two quadratures have complementary phases and
their quantum noise is not correlated. At a time $\tau=0.5$, one
phase is changed so the quadratures become anti-correlated.\label{fig:Graph-of-xp-correlations}}}
\end{figure}
 Noting that the LO interaction with the output field is reversible,
the setting of system $1$ is then changed from $\theta_{1}=0$ to
$\theta_{1}=\pi/2$, so that $\hat{p}_{1}$ would be measured directly.
The value of $J_{2}(\tau)$ is not changed by the change in setting
at system $1$ (no-signaling) and the stochastic currents $J_{2,av}(\tau)$
and $J_{1,av}(\tau)$ become anti-correlated (Fig. \ref{fig:Graph-of-xp-correlations}).
At the time $t_{D2}$, the value of the stochastic current $J_{2,av}(\tau)$
gives the correct prediction $p_{1}=-p_{2}$ for the outcome of $\hat{p}_{1}$,
should that measurement be performed at system $1$. Hence, we can
say that the element of reality (EOR) for system $1$ is valid at
time $t_{D2}$, when the setting at system $2$ has been finalized.

The argument of EPR posits further that the EOR for system $1$ exists,
whether or not the measurement setting $\theta_{2}$ has been finalized
at system $2$, since according to their premises, that procedure
in no way disturbs system $1$. This implies that the EOR exists prior
to the choice of both settings $\theta_{1}$ and $\theta_{2}$. This
stronger assumption applies when both systems are microscopic, and
is not required in our work. Further details are in Appendix F.

\paragraph*{Two mode Ising problem}

To illustrate an application to quantum technology, the homodyne currents
from solving an SSE can also be used to identify the Ising problem
ground state spin configurations. This is a hard computational problem
solved by a coherent Ising machine (CIM) \citep{wang2013coherent,Marandi2014,yamamoto2017coherent,McMahon_CIM_science2019,Yamamoto2020,Thenabadu_arxiv2025},
which outputs homodyne currents to compute spin configurations. Here,
the device is taken to be in an initial vacuum state, evolving to
a final state which encodes the target solution, where each mode is
pumped with an identical amplitude $\lambda$, described by a Hamiltonian
$H_{p}=i\hbar\lambda/2\sum_{j}\left(a_{j}^{\dagger2}-a_{j}^{2}\right)$.

A positive (negative) quadrature output is mapped to an Ising spin
$\sigma$ of $+1$ (-1), and the Ising energy for a set of spin configurations
$E(\bm{\sigma})=-\sum_{kj}C_{kj}\sigma_{k}\sigma_{j}$. To obtain
a quantum advantage, it is essential to operate these devices in highly
nonlinear regimes described by a master equation \citep{Thenabadu_arxiv2025}
rather than the classical equations applicable at low nonlinearity.
This nonlinearity appears in the master equation as damping operators
$a_{j}^{2}$, with corresponding nonlinear decay $g^{2}/2$.

\begin{figure}
\begin{centering}
\includegraphics[width=0.5\columnwidth]{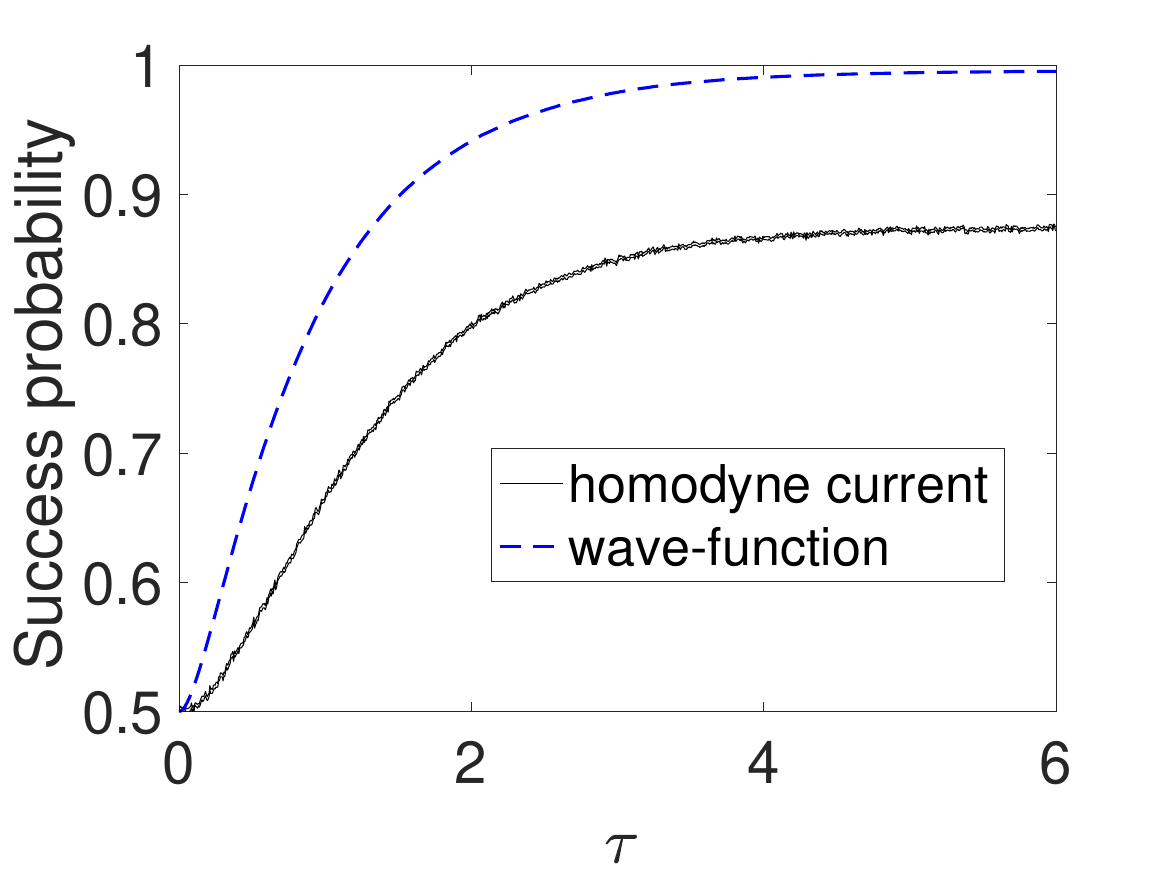}\includegraphics[width=0.5\columnwidth]{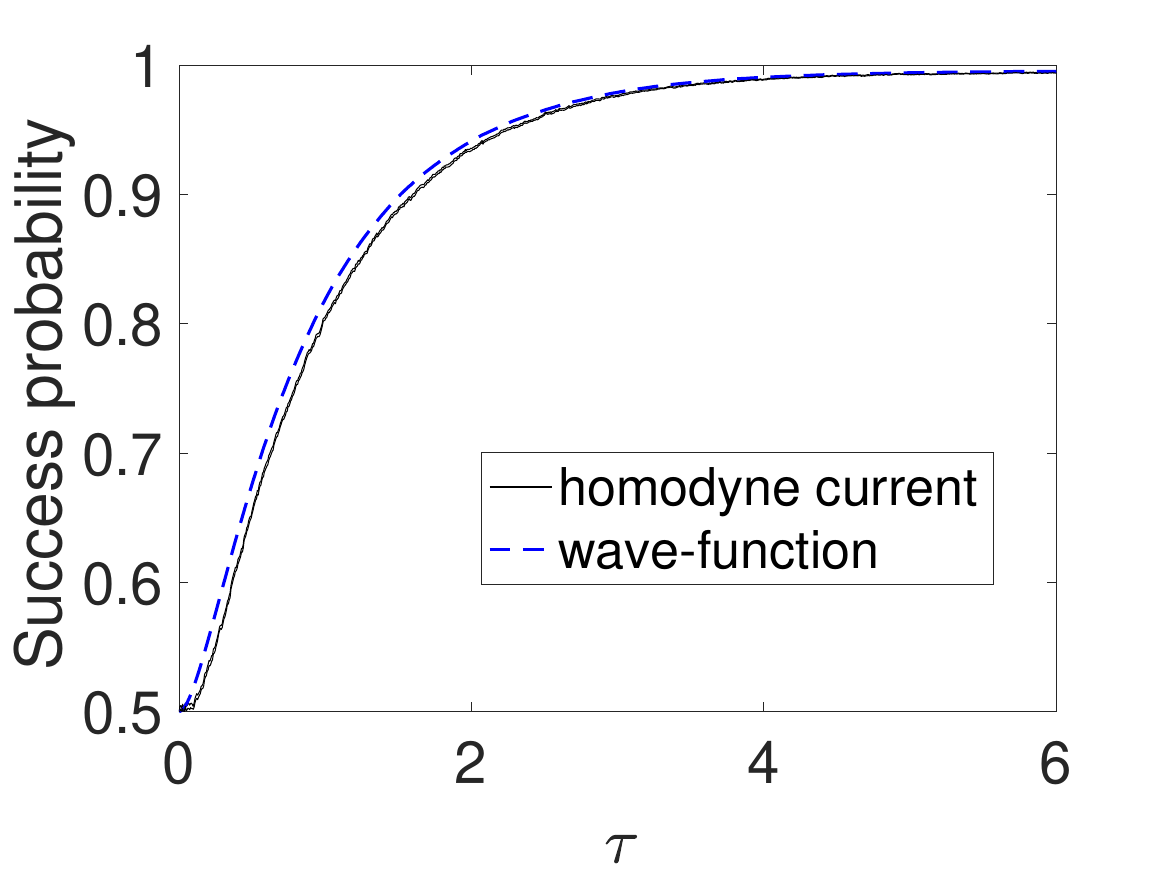}
\par\end{centering}
\caption{The two-mode CIM success probability inferred from the filtered homodyne
current outputs (black line) and from the wave-function (red dashed
line). Both results are derived from solving the same homodyne SSE.
Left graph, $\kappa=50$, right graph, $\kappa=10$, using $10^{4}$trajectories
and a time-step of $\Delta\tau=0.003$. \label{fig:TSuc_prob_filtered_kappa50}}
\end{figure}

To illustrate this, we take a simple two mode ferromagnetic Ising
problem with a coupling matrix $\bm{C}=\begin{pmatrix}0 & 1\\
1 & 0
\end{pmatrix}$, where the lowest energy states have spin configurations with aligned
spins, to show how homodyne noise can affect results in a deeply quantum
regime. We evaluate how well the machine performs by the success probability
of it finding the correct ground state spin configuration. Relevant
parameters are the pump amplitude $\lambda=2.4$, and a nonlinear
decay parameter $g=0.6$.

Surprisingly, this example shows that the noise from a high-bandwidth
filter cannot be removed by using more trajectories and averaging.
The quadrature sign is a discontinuous and nonlinear function, which
introduces systematic errors that are still present after averaging. 

The success probability time evolution inferred from the filtered
homodyne currents is presented in Fig. \ref{fig:TSuc_prob_filtered_kappa50}.
In the figure, the success probability is computed in two ways: one
is inferred from the filtered homodyne current, and the other is calculated
from the SSE wave-function, using standard quantum measurement theory.
The results do not agree because the SSE includes a more realistic
shot-noise model.

The agreement depends on the precise detection bandwidth chosen. In
Fig. \ref{fig:TSuc_prob_filtered_kappa50}, the detection bandwidth
$\kappa$ is taken to be $5$, which is $5$ times the single photon
decay rate. When $\kappa=50$ is chosen instead, the success probability
from these two methods no longer agrees. The increased noise bandwidth
allows the white noise at the detector to change the sign of the measured
quadrature. This shows that a narrowband filter is essential for operation
in a highly quantum regime.

\paragraph*{Summary}

In conclusion, we have shown that the EPR argument can be used to
identify SSE outputs representing measured currents. Our conclusion
is that the Stratonovich form of SSE current corresponds to a physical
current, in the wide-band limit. More realistically, one should use
a finite bandwidth model. This also gives information on systematic
errors in quantum technologies and innovative tests of quantum foundations.

\paragraph*{Acknowledgements: }

This publication was made possible through an NTT Phi Laboratories
grant, and support of Grant 62843 from the John Templeton Foundation.
The opinions expressed in this publication are those of the author(s)
and do not necessarily reflect the views of the John Templeton Foundation.

\subsection*{Data Availability Statement}

All simulations were performed using the publicly available software
package, xSPDE \citep{Drummond2025xspde4} .

\bibliographystyle{apsrev4-2}

\pagebreak{}

\newpage{}

\section*{End Matter}

\paragraph*{Appendix A: Ito and Stratonovich SSE}

We start with the homodyne current SSE (\ref{eq:Ito_SSE}) in the
Ito calculus \citep{carmichael2009statistical}. This Ito stochastic
differential equation (SDE) can be written in terms of its number
state expansion as: 
\begin{equation}
\frac{\ensuremath{d\psi_{k}}}{d\tau}=A_{k}^{(I)}+B_{kj}\xi_{j}
\end{equation}
Correction terms have to be worked out to give the corresponding Stratonovich
SDE, which will have different drift terms $A$. These terms are calculated
from an expression \citep{Gardiner1997} that holds for each component
$\psi_{j}$ of the conditional wavefunction
\begin{align}
A_{i} & =A_{i}^{(I)}-\frac{1}{2}\sum_{kj}B_{jk}\frac{\partial}{\partial\psi_{j}}B_{ik}-\frac{1}{2}\sum_{kj}B_{jk}^{*}\frac{\partial}{\partial\psi_{j}^{*}}B_{ik}\,.
\end{align}
After carrying out the differentiations, the result in the main text
is obtained.

\paragraph*{Appendix B: The generating functional method}

How did previous theoretical approaches identify a model for the detected
homodyne current, and what was the physics issue? To summarize \citep{Barchielli_1986,gardiner2004quantum},
one defines a generating functional both in quantum mechanics and
in SSE theory, which is used to calculate all measurable correlations
and moments, where:
\begin{equation}
\Phi_{t}[\bm{k}]=\left\langle \exp\left\{ \int_{0}^{t}\bm{k}(s)\cdot d\bm{X}_{out}\left(s\right)\right\} \right\rangle .
\end{equation}

If the generating functional has a stochastic differential equation
that is identical for quantum theory and the SSE current, then the
generating functionals are the same. The simulated current would therefore
be realistic, since this defines a unique probability. However, an
Ito stochastic equation for the SSE generating functional requires
correction terms that depend on the equations \citep{gardiner2004quantum}.
Such terms cannot occur for any physical current which is continuous,
even as a limit.

Therefore, the generating functional argument cannot be used to identify
a current prior to filtering, if it includes Ito noises that require
correction terms in the proof. However, the argument is valid for
a Stratonovich noise term which has no corrections, and uses the same
calculus as a physical current. This is the foundational question
of how to identify a realistic current model in the context of SSE
quantum theory.

The results given above using adiabatic elimination, as well as the
equal-time EPR correlation example agree with this logic. Therefore,
if one wishes to obtain a model for the homodyne current as an element
of reality prior to filtering, the Stratonovich model of the added
shot noise is the only suitable candidate. 

\paragraph*{Appendix C: Stratonovich current correlations}

We now show how the quantum operator result $\langle\hat{J}_{1}(\tau)\hat{J}_{2}(\tau)\rangle_{Q}$
is related to the current trajectory choice, using both a short-time
solution and a full numerical integration. To achieve this we compare
Ito calculus, in which the noise and the wave-function are defined
at the start of a time interval, and are not correlated \citep{ito1951formula},
with Stratonovich calculus, in which the noise and the wave-function
are defined at the center of a time interval. Our goal is to find
a stochastic current model that agrees with the quantum predictions,
ie, $\langle\hat{J}_{1}(\tau)\hat{J}_{2}(\tau)\rangle_{Q}=\left\langle J_{1}(\tau)J_{2}(\tau)\right\rangle $,
where $\left\langle .\right\rangle $ indicates an ensemble average
over the stochastic trajectories. Therefore, for Stratonovich noise
we first treat an analytic short-time solution in an interval from
$\tau=0$ up to $\tau=\Delta\tau$, with the noise evaluated at the
midpoint, ie, $\tau=\bar{\tau}=\Delta\tau/2$. The noise can be treated
as constant on a short interval \citep{stratonovich1971probability,drummond1991computer},
and the wide-band limit corresponds to $\Delta\tau\rightarrow0$.

If we take a finite but short time-step, then the short-time current
is dominated by the noise term in Eq. (\ref{eq:Stratonovich_noise}),
scaling as $1/\sqrt{\Delta\tau}$. To show this, consider a random
noise $\Delta\bm{w}$ with ${\normalcolor \langle\left(\Delta w_{k}\right)^{2}\rangle}=\Delta\tau$.
The average noise term is $\bar{\bm{\xi}}(\bar{\tau})=\Delta\bm{w}/\Delta\tau.$
The leading term in either current at time $\bar{\tau}$ is of order
$\Delta\tau^{-1/2}$, and comes from the broad-band noise, where for
Stratonovich noise:
\begin{equation}
\bm{J}(\bar{\tau})=\langle\psi(\bar{\tau})|\hat{\bm{x}}|\psi(\bar{\tau})\rangle_{S}+\bar{\bm{\xi}}_{S}\left(\bar{\tau}\right).
\end{equation}

Noise terms will average to zero unless multiplied by a correlated
noise term which scales as $\sqrt{\Delta\tau}$, giving a term of
$O(1)$. Since $\left\langle a_{k}(0)\right\rangle _{c}=0$, to order
$\sqrt{\Delta\tau}$ the conditional wave-function is given by Eq.
(\ref{eq:Strat_SSE}) as:
\begin{equation}
|\psi(\bar{\tau})\rangle_{S}=\left(1+\frac{1}{2}\hat{\bm{a}}\cdot\Delta\bm{w}+O\left(\Delta\tau\right)\right)|\psi(0)\rangle.
\end{equation}
The leading term in $\left\langle \hat{x}_{k}(\bar{\tau})\right\rangle _{S}$
has the required scaling of $O(1)$, as it is conditioned by a complementary
noise:
\begin{align}
\left\langle \hat{\bm{x}}(\bar{\tau})\right\rangle _{S} & =\frac{1}{2}\sum_{k}\langle\psi(0)|\Delta w_{k}\left(\hat{a}_{k}^{\dagger}\hat{\bm{x}}+\hat{\bm{x}}\hat{a}_{k}\right)|\psi(0)\rangle.
\end{align}

Since we wish to compute $\langle J_{1}(\Delta\tau)J_{2}(\Delta\tau)\rangle$,
the only significant term which gives a non-vanishing noise correlation
is for $i=3-k$. Accordingly, only keeping terms with $i\neq k$,
one has that
\begin{equation}
\left\langle x_{k}(\bar{\tau})\right\rangle _{S}=\frac{1}{2}\Delta w_{3-k}\left\langle \hat{x}_{3-k}\hat{x}_{k}\right\rangle +O\left(\Delta w_{k}\right)+O\left(\Delta\tau\right).
\end{equation}
There are two equal terms in the current correlation, each of which
is a correlation between a conditional expectation value and a noise
term in the complementary field, giving the correct result, independent
of $\Delta\tau$ in the short time-step limit:
\begin{equation}
\lim_{\tau\rightarrow0}\left\langle J_{1}(\tau)J_{2}(\tau)\right\rangle =\left\langle \hat{x}_{1}\left(0\right)\hat{x}_{2}\left(0\right)\right\rangle _{Q}.
\end{equation}
This is in agreement with input-output theory at short times. To treat
longer times, we have solved the homodyne wide-band Stratonovich SSE,
using a midpoint algorithm \citep{drummond1991computer} and public
domain quantum SDE software, xSPDE \citep{Drummond2025xspde4}. Choosing
a squeezing parameter $r=0.5$, we have computed $\left\langle J_{1}(\tau)J_{2}(\tau)\right\rangle $
for a finite ensemble, obtaining results identical to those from the
exact quantum solutions up to sampling errors, as shown in Fig. \ref{fig:j1j2_unfiltered}.
Apart from sampling errors, this is also independent of time-step.

\paragraph*{Appendix D: Ito current correlations}

One can carry out this calculation with an Ito noise current, by evaluating
the noise at the same time as the wave-function. In the formalism
of Ito calculus, these are uncorrelated, and so $\langle j_{1}^{I}(\tau)j_{2}^{I}(\tau)\rangle=\left\langle \left\langle \hat{x}_{1}(\tau)\right\rangle _{I}\left\langle \hat{x}_{2}(\tau)\right\rangle _{I}\right\rangle $.
This implies that $\lim_{\tau\rightarrow0}\langle j_{1}^{I}(\tau)j_{2}^{I}(\tau)\rangle=0$,
agreeing with the result in Fig. (\ref{fig:j1j2_unfiltered}) which
does not give the correct quantum correlations. This result was obtained
both with xSPDE and with another quantum software package, Qutip \citep{qutip1}.

As an alternative, it is sometimes proposed \citep{Gambetta_PRA2002}
that one must use the Ito noise term in the interval preceding the
time of the wave function. However, this gives too strong an initial
correlation of $\lim_{\tau\rightarrow0}\langle j_{1}^{d}(\tau)j_{2}^{d}(\tau)\rangle=2\left\langle \hat{x}_{1}\left(0\right)\hat{x}_{2}\left(0\right)\right\rangle _{Q}$,
which is also incorrect, as shown in Fig. (\ref{fig:j1j2_unfiltered}).
This verifies our result that the wide-band homodyne current cannot
be the Ito based current $\bm{j}^{I}(\tau)$ or $\bm{j}^{d}(\tau)$.

\paragraph*{Appendix E: Effects of finite bandwidth}

\begin{figure}
\begin{centering}
\includegraphics[width=0.5\columnwidth]{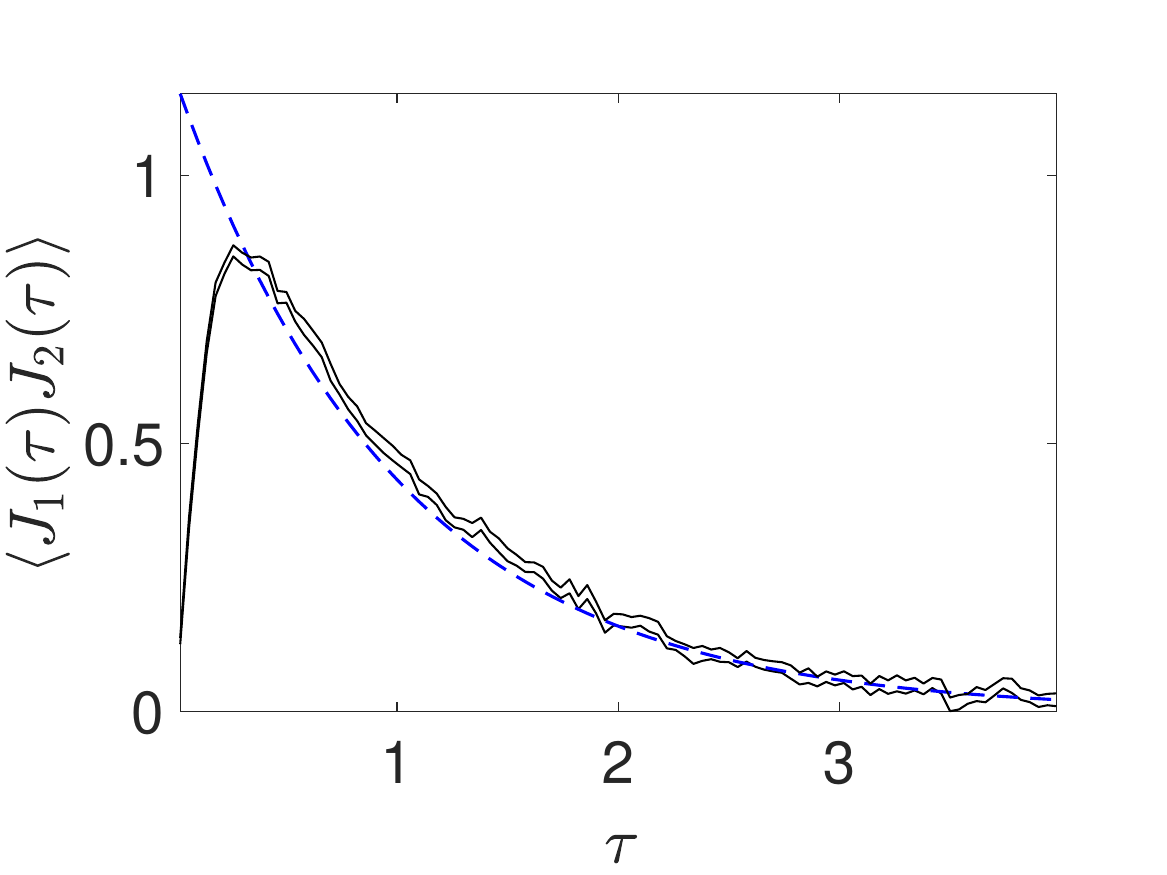}\includegraphics[width=0.5\columnwidth]{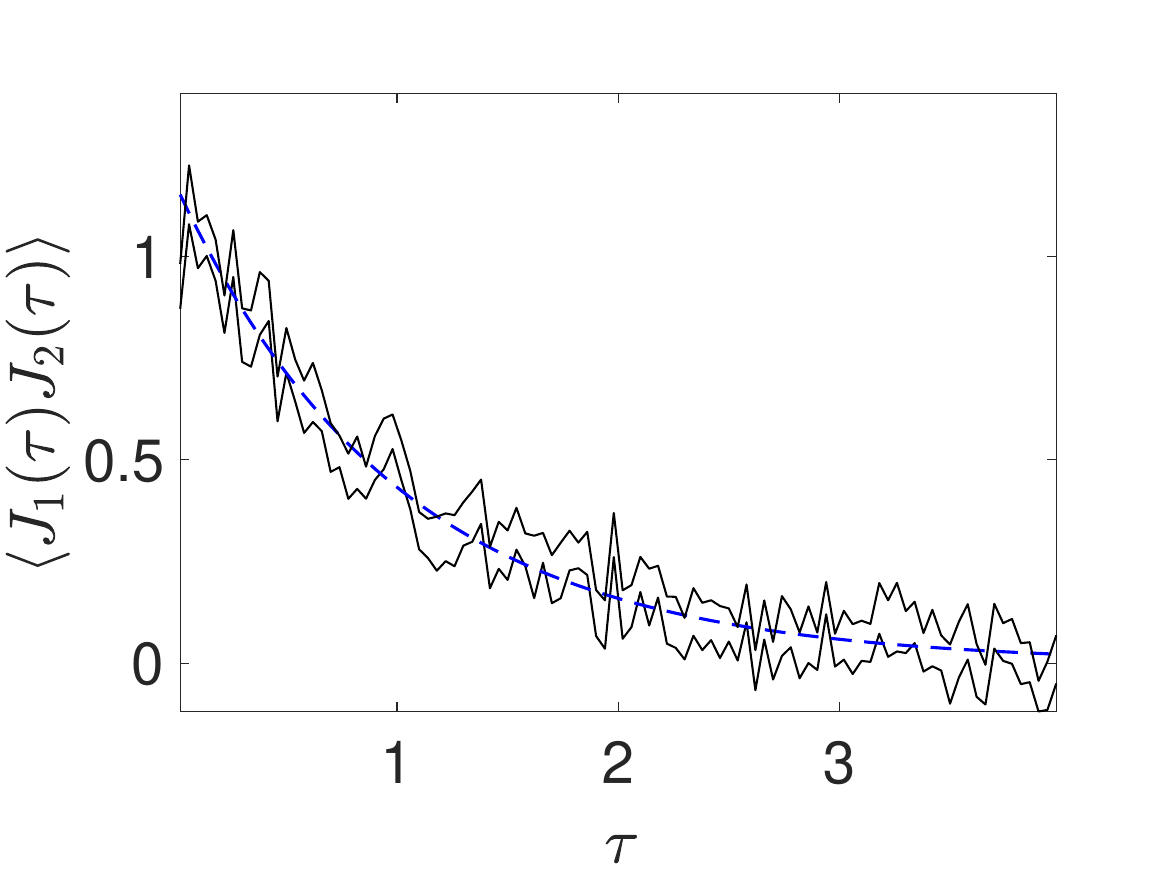}
\par\end{centering}
\caption{The averaged filtered homodyne current correlation $\langle J_{1}(\tau)J_{2}(\tau)\rangle$
vs the dimensionless time $\tau=\gamma t$ for a two-mode damped squeezed
state with different detection bandwidth $\kappa$. The left and right
plots correspond to detection bandwidths $\kappa$ of $10$ and $50$
respectively. The blue dashed line is the infinite bandwidth analytical
solution, $\langle J_{1}(\tau)J_{2}(\tau)\rangle=-e^{-\tau}\sinh(2r)$.
The noise in the current and the wave-function are evaluated at the
same time. Here, $r=0.5$ is the squeezing parameter, the two lines
are sampling error bars for a sample size of $2\times10^{5}$, and
the time step-size is $0.001$. \label{fig:j1j2_filtered}}
\end{figure}

A finite-bandwidth current is an even better representation of a physical
measurement, where the detection bandwidth changes the signal. The
bandwidth $\kappa$ gives the range of frequencies over which a detector
can faithfully respond to a signal. A high bandwidth with respect
to the damping rate resolves rapid changes in the signal, tracking
the dynamics well, as demonstrated in Fig. \ref{fig:j1j2_filtered}
with $\kappa=50$.

This also generates a noisy signal due to detector shot-noise. In
comparison, a bandwidth of $\kappa=10$ produces a signal which does
not track the dynamics as closely, but has less noise. These results
reveal a trade-off between well-resolved dynamics and noise. To track
the dynamics accurately at high bandwidth, we must take a larger number
of samples to reduce the output noise in the signal, while a reduced
bandwidth requires fewer samples.

In the time-averaged limit, one can also use spectral methods \citep{drummond2004quantum}
to analyse squeezing, or a mode-matched, pulsed local oscillator \citep{Slusher1987pulsed,Drummond1993-solitons}.
After integration over all times these give correlations without excess
noise, but also without any dynamical information.

\paragraph*{Appendix F: Changing the phase-angle;}

In the gedanken-experiment we describe here, it is possible to change
measurement settings dynamically, so that the modified premises can
be directly tested. The measurement settings correspond to phase-angles
of the local oscillator. Similar dynamical experiments have been carried
out, for example a superconducting experiment on reversing quantum
jumps. although here there was only a single observable \citep{minev2019catch}.
A related, although not time-resolved, EPR experiment was recently
reported using macroscopic Bose-Einstein EPR correlations \citep{colciaghi2023einstein}.

To illustrate how the local phase-angle is relevant, we simulate an
experiment in which the phase-angle at one meter is fixed, while it
is varied at the other. This is similar conceptually to experiments
carried out with superconducting quantum devices in which a quantum
jump is interrupted in midflight \citep{minev2019catch}. By extending
to two modes, we illustrate how one could carry out the original proposal
of Schr{\"o}dinger \citep{schrodinger1935gegenwartige}, in his response
to EPR's argument.

In Fig. (\ref{fig:Graph-of-xp-correlations}), the first meter is
set to measure an $x$-quadrature, while the second meter first measures
a $p$- quadrature, then an $x$-quadrature. As expected from quantum
mechanics, the output current element of reality, and hence the correlations
depend on the phase-setting. This illustrates Schr{\"o}dinger's original
measurement proposal. In the correlated state, an $x_{1}$ variable
is measured at one location, and is correlated with an $x_{2}$ variable
at another location.

At the second location, one can choose to either measure $x_{2}$
or $p_{2}$. The case that one measures $p_{2}$, while also measuring
$x_{1}$, with the settings adjusted at a time $t_{?}$, creates the
paradox that one apparently has a knowledge of both $x_{2}$, through
its anti-correlation with $x_{1}$, and $p_{2}$, even though they
cannot be measured simultaneously. This represents an alternative
EPR argument for the incompleteness of quantum mechanics, based on
modified premises that are not negated by Bell's theorem. In Fig.
(\ref{fig:Graph-of-xp-correlations}) we confirm the validity of the
``element of reality'' for $x_{2}$, by carrying out the change
of setting at system $2$, from $p_{2}$ to $x_{2}$, and showing
that the final value $x_{2}$ is correlated with $x_{1}$, at the
level required for an EPR criterion. 
\end{document}